\documentclass{elsart}
 
\usepackage{amssymb}
\usepackage{graphicx}
\usepackage[english,francais]{babel}

\newtheorem{e-proposition}[theorem]{Proposition}

\newtheorem{e-definition}[theorem]{Definition\rm}

\def\og{\leavevmode\raise.3ex\hbox{$\scriptscriptstyle\langle\!\langle$~}}
\def\fg{\leavevmode\raise.3ex\hbox{~$\!\scriptscriptstyle\,\rangle\!\rangle$}}

\begin{document}
\centerline{GRBs and the high-redshift Universe $-$ Sursauts Gamma et l'univers lointain}
\begin{frontmatter}


\selectlanguage{english}
\title{Gamma Ray Bursts as Probes of the Distant Universe}


\selectlanguage{english}
\author[authorlabel1]{Patrick Petitjean}
\ead{petitjean@iap.fr}
\&
\author[authorlabel2]{Susanna D. Vergani}
\ead{susanna.vergani@brera.inaf.it}

\address[authorlabel1]{Universit\'e Paris 6, UMR7095, Institut d'Astrophysique de Paris, CNRS\\
98bis Boulevard Arago, 75014 Paris $-$ France}
\address[authorlabel2]{INAF - Osservatorio Astronomico di Brera, Via Bianchi 46, I$-$23807 Merate, Italy}

\begin{abstract}
We review recent results on the high-redshift universe and the cosmic evolution
obtained using Gamma Ray Bursts (GRBs) as tracers of high-redshift galaxies. Most of the results come from
photometric and spectroscopic observations of GRB host galaxies once the afterglow has faded away but also
from the analysis of the GRB afterglow line of sight as revealed by absorptions in their optical spectrum. 

\vskip 0.5\baselineskip

\selectlanguage{francais}
\noindent{\bf R\'esum\'e}
\vskip 0.5\baselineskip
\noindent
{\bf Comment sonder l'Univers lointain \`a l'aide des sursauts gamma. }
Nous passons en revue les r\'esultats obtenus r\'ecemment sur les propri\'et\'es de l'univers \`a grand d\'ecalage
spectral et l'\'evolution cosmique, en utilisant les sursauts gamma comme des traceurs
de galaxies lointaines. La plupart des r\'esultats viennent de l'observation en imagerie et spectroscopie des galaxies
h\^otes des sursauts, une fois que le sursaut s'est \'eteint, mais aussi de l'analyse des absorptions observ\'ees dans le spectre de la r\'emanence.

\keyword{GRBs, cosmology, host-galaxy, absorption lines, interstellar medium}  \vskip 0.5\baselineskip
\noindent{\small{\it Mots-cl\'es~:} Sursauts gamma; cosmologie; galaxie h\^ote; raies d'absorption; milieu inter-stellaire}}
\end{abstract}
\end{frontmatter}

\selectlanguage{english}

\section{Introduction}

\label{}
Gamma-ray bursts (hereafter GRBs) are much more than powerful flashes of high-energy photons 
that travel undisturbed from cosmological distances to Earth. They are unique probes of the 
high-redshift Universe and of the cosmic evolution.

GRBs flag remote galaxies that would probably stay unnoticed otherwise
(e.g. \cite{Chen2009}). They are divided in two classes: `short' and `long' GRBs, depending to a first approximation 
on the duration of their prompt emission. At least a fraction of long GRBs have    
been shown to be associated with the collapse of massive stars via the observation of supernova 
signatures in the light curves once the GRB optical afterglow has faded away (e.g. \cite{Hjorth2003}\cite{Stanek2003}). 
The long GRB-supernova connection implies that GRBs probably track the formation of massive stars and could be
used as a complementary probe of the global star formation rate (SFR) in galaxies in the high-$z$ regime
where data are lacking. In addition, GRBs must happen in the primordial Universe in association with
the death of the first stars and are therefore intimately related to the reionization of the Universe
(e.g. \cite{Bromm2006}).

GRB afterglows are, but only for a small spell of time, bright background sources in the spectrum of 
which the objects located along the line of sight imprint their shadows in the form of specific absorption
lines (e.g. \cite{Paul2007}). They will be particularly interesting when it will be possible to
detect and observe them at very high redshifts
where their line of sight will reveal the state of the intergalactic medium during the dark ages.
On the other hand, absorption signatures from the gas belonging to the host galaxy itself
are revealed (e.g. \cite{Savaglio2006}). GRB host-galaxies are therefore unique objects where both 
emission and absorption from the interstellar medium can be observed.  

Finally they might also be important to constrain the small-scale power spectrum of primordial density
fluctuations \cite{Mesinger2005}.


\section{GRBs and the cosmic star formation rate}
\label{zdist}
Although convincing evidence in favour of the association of long GRBs 
(e.g. GRBs with a duration $T_{\rm 90}$$>$2~s) with massive stars has been gathered 
over the recent years \cite{Stanek2003}\cite{Hjorth2003}, the actual physical conditions (e.g. mass, metallicity, 
rotation, binarity) for a star to trigger a burst are not currently known (see however \cite{Yoon2006}\cite{Woosley2006}). 
Thus, although it is expected that the rate of GRBs in the Universe and star formation rate are related, there must be 
so in a non-trivial way. 

\subsection{Comparison of the star formation and GRB rates}

The NASA {\sl Swift} satellite, launched in 2004 November, heralded a new era of 
rapid GRB localization. X-ray and UV telescopes on board {\sl Swift} provide the means to localize                            
GRBs with small error boxes, so that dedicated ground-based telescopes can image the 
fading optical afterglow.  The synergy between {\sl Swift}'s sensitivity and localization capabilities 
and the growing number of rapid response ground-based telescopes has greatly improved the number of 
GRB redshifts that could be determined. Optical/NIR afterglows have been found            
for about 42\% of GRBs, and 62\% of these GRBs with an optical afterglow have measured redshifts 
(see http://www.mpe.mpg.de/$\sim$jcg/grbgen.html). 
Although the mean redshift for {\sl Swift} afterglows ($z\sim 2$) is larger than that of
pr\'e-{\sl Swift} afterglows ($z\sim 1.1$)
it seems that shorter response times favour optically fainter bursts that are relatively closer
\cite{Coward2009}. This selection effect could explain why the average redshift for
{\sl Swift} afterglows, $z\sim$2.8 as measured in 2005, has evolved to $z\sim2$. 

The distribution of detected GRBs is given in the left panel of Fig.~\ref{fig1}. Only 
GRBs with a sufficient luminosity have been considered, 
$L_{\rm iso}$~=~$E_{\rm iso}$/[$T_{90}$/(1+$z$)]~$\\>$~10$^{51}$~erg/s,
with $E_{\rm iso}$ the rest frame isotropic 1$-$10$^4$~keV energy release
and $T_{90}$ the time interval containing 90\% of the prompt emission. 
On the figure, the shaded region illustrate the chosen threshold.
Details of the samples can be found in \cite{Kistler2009} and \cite{Butler2007}.

\subsection{Extrapolation to the highest redshifts}

\begin{figure}
\centering
\hbox{
\includegraphics[width=0.49\linewidth, angle=0]{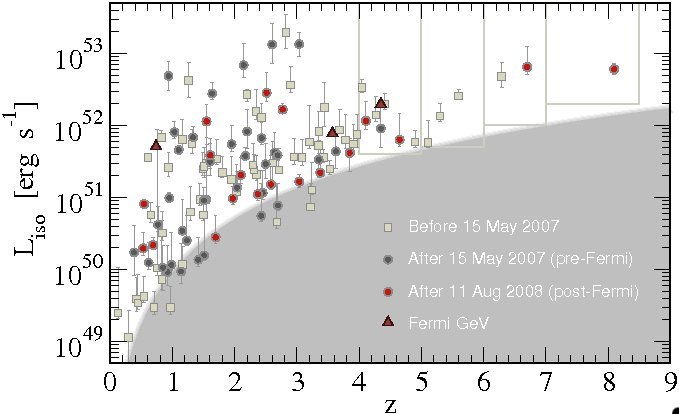}
\includegraphics[width=0.49\linewidth, angle=0]{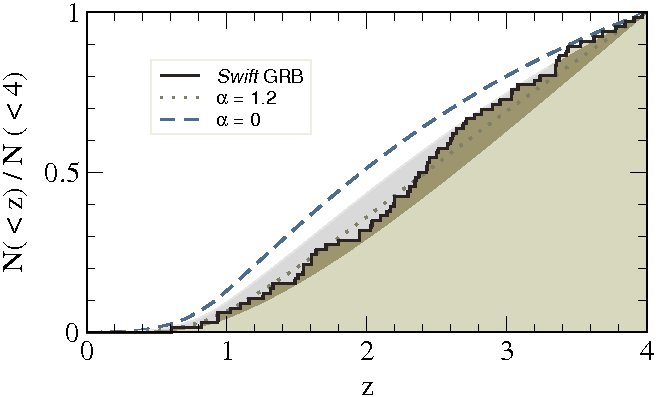}
}
\caption{{\sl Left panel:} GRB luminosity versus redshift (for GRBs with $T_{\rm 90}>2$~s) 
as to June 2009 (see \cite{Kistler2009} for details). The shaded region approximates an effective threshold
of detection. {\sl Right panel:} The cumulative distribution of {\sl Swift} GRBs with 
$L_{\rm iso}$~$>$~10$^{51}$~erg/s shown as the solid histogram is compared with the
cosmic star formation rate from \cite{Hopkins2006} (scaled as the long-dashed line). The latter must be corrected
by a factor (1~+~$z$)$^{\alpha}$ with $\alpha$~=~1.2 to fit the GRB rate. The two shaded 
regions on both sides are for $\alpha$~=~0.6 and 1.8. This curve is used to extrapolate the cosmic star 
formation rate to higher redshifts from the GRB rate. The figure is taken from \cite{Kistler2009}.
}
\label{fig1}
\end{figure}

In the righ panel of Fig.~\ref{fig1}, the cumulative GRB rate is shown versus redshift as an histogram when 
the cosmic SFR is scaled as a long-dashed line. It is apparent that both rates do not correlate well
(see also \cite{Daigne2006}\cite{Porciani2001}).
This can be a consequence of (i) the GRB detection and redshift determination being biased by observational 
systematics (e.g. \cite{Imerito2009}\cite{Jakobsson2009}\cite{Fynbo2009}), (ii) the complexity of the relation
between GRB rate and star formation rate (\cite{Rossi2007}\cite{Svensson2010}) e.g. long GRBs are
associated with SNIc which have not been demonstrated to be related to general star formation activity.


To calibrate the differential evolution between the GRB rate and the SFR, the latter can be multiplied by a factor 
(1~+~$z$)$^{\alpha}$. The index is fitted so that the two rates match. A good fit is obtained with 
$\alpha$~=~1.2 (see Fig.~\ref{fig1}).
Assuming the correction holds to redshifts higher than $z=4$, the SFR can be extrapolated using the GRB rate to beyond
this redshift where optical data are lacking. Although the extrapolation may be hazardous in particular
because GRB progenitors may not form in the typical star formation environment throughout all
redshifts, the exercise may still be worth because GRBs should be detectable at any redshift 
\cite{Tanvir2007}\cite{Salvaterra2008} when objects at $z>7$ are very difficult to detect (e.g. \cite{Bouwens2010}). 
Actually the GRB with highest redshift up to now has $z=8.26$ \cite{Tanvir2009}\cite{Salvaterra2009}.

\section{The host-galaxies of GRBs}

\subsection{Long GRBs}

The study of GRB host galaxies is very important to the understanding of the physical 
properties of regions where GRBs explode and, as a consequence, the nature of GRB progenitors. Although a 
large number of host galaxies can be identified with deep imaging, most of them are too faint 
to be observed even using the largest telescopes and so far, only a few dozens of host galaxies have 
been observed spectroscopically. Moreover, the number of GRB hosts that have been intensively 
studied is even less. In the most extensive study of GRB hosts, Savaglio et al. \cite{Savaglio2009}
study the properties  of 46 GRB host galaxies over a redshift interval  of $0 < z < 6.3$ 
(for a median redshift of $z = 0.96$ corresponding to a look-back time of 7.2~Gyr),  
most of them (89\%)  being at $z<1.6$. The sample of \cite{Chen2009} includes 15 objects at $z>2$.
It must be noted that it may be very dangerous to derive detailed
conclusions from so small samples covering the whole Universe history especially with so
few information during the time of highest star formation rate ($z>1$; \cite{Hopkins2006}).

Photometric and spectroscopic observations show that the host galaxies of 
long-duration GRBs are mostly faint, blue, low-mass, star-forming galaxies with low metallicities 
(e.g. \cite{LeFloch2003}\cite{Prochaska2004}\cite{Christensen2004}\cite{Fynbo2005}\cite{Gorosabel2005} 
\cite{Wiersema2007}\cite{Savaglio2009}\cite{Han2010}. In contrast, the
host galaxies of short-duration GRBs mostly have higher luminosities and 
higher metallicities than long-duration GRB hosts \cite{Leibler2010}.


\subsubsection{The low-$z$ view; $z<1.5$}

Since the conditions in the Universe are very different through its cosmic evolution,
the formation of GRBs could be very different at low and high redshift. It is
easier however to gather information at low redshift especially for the GRB host-galaxies.

Svensson et al. \cite{Svensson2010} compare the photometric properties and small scale
environments of 34 GRBs and 58 core colapsed supernovae (CCSN) host galaxies at $z<1.2$ observed
with {\sl HST}. They find that while GRB hosts are typically both smaller and bluer
than those of CCSN, their total blue light luminosities are only slightly lower possibly because
of rapid periods of intensified star formation activity which both create GRB progenitors and
briefly significantly enhance the host galaxy blue luminosity.

\begin{figure}
\centering
\includegraphics[width=7.5cm, angle=0]{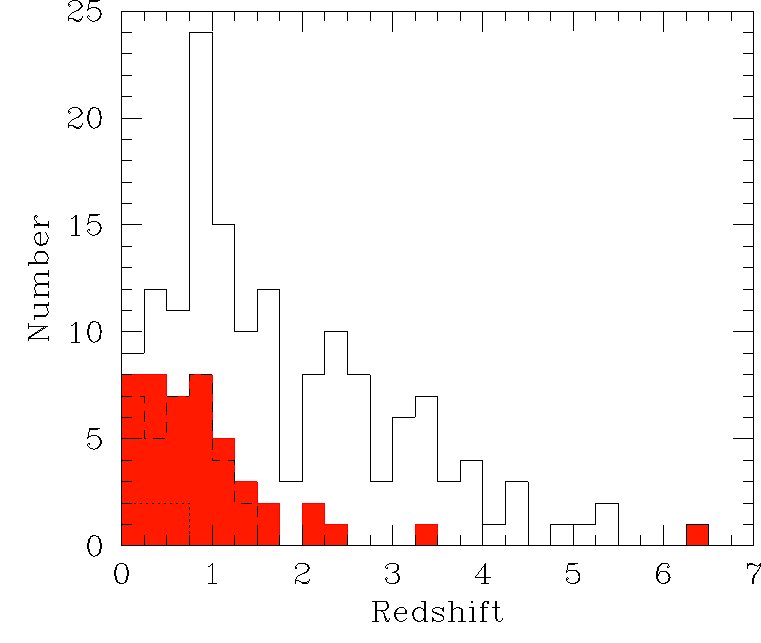}
\caption{Redshift distribution in a sample of GRBs (solid black histogram) and
the corresponding host-galaxies (red shaded regions) from \cite{Savaglio2009}. It is apparent that host-galaxies
are difficult to detect at high redshift ($z>2$) and that most of the information gathered
on host galaxies apply to $z<1.5$ (see however \cite{Chen2009}). The figure is from \cite{Savaglio2009}.
}
\label{Distzhost}
\end{figure}

The distribution of redshifts of the host galaxies studied spectroscopically by \cite{Savaglio2009} is shown
in Fig.~\ref{Distzhost} together with the distribution of GRB redshifts. It is apparent
on the Figure that host-galaxies are difficult to detect at $z>2$ and that most
of the information is from galaxies with $z<1$.

\bigskip
$\bullet$ {\sl Stellar mass and star formation}

\begin{figure}
\centering
\includegraphics[width=8cm, height=6.5cm, angle=0]{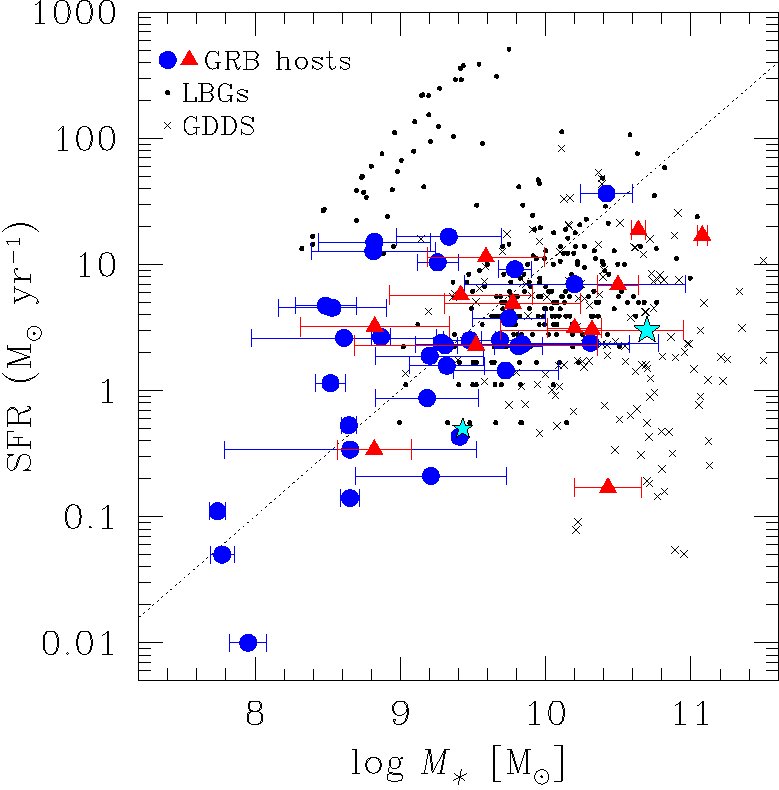}
\caption{Star formation rate (SFR) versus stellar mass ($M_{\rm *}$) of GRB host
galaxies (blue and red points) compared to measurements for Lyman break galaxies (LBGs, \cite{Reddy2006}) and 
field galaxies observed by the Gemini Deep Deep Survey (GDDS, \cite{Glazebrook2004}).
Blue circles are for SFRs measured using H$\alpha$ and [OII] and red circles are for
SFRs measured from UV 2800~\AA~ luminosities.
The figure is from \cite{Savaglio2009}.
}
\label{SFRM}
\end{figure}

Stellar population synthesis models are used to fit the SED of galaxy hosts to derive
the main physical parameters of the galaxy (stellar mass and age, metallicity, extinction, characteristics
of the burst component etc..). Such a model has been consistently applied to the SED of
host-galaxies in the sample of \cite{Savaglio2009}. The model has two components, one representing
an old population of star formation history varied from an old starburst to a constant
SFR, and a second component representing a young population originating in a recent burst.
The mass of the young component can be anywhere from 10$^{-4}$ to twice the mass
of the old component (see \cite{Savaglio2009} for details).
Stellar mass is derived from the infra-red colors (see \cite{Glazebrook2004}), SFR is
derived from the UV flux and the emission lines. 
Results from this sample is shown in Fig.~\ref{SFRM}.
The mean stellar mass is similar to the stellar mass of the Large Magellanic Cloud (LMC), 
$M_{\rm *}$~$\sim$~10$^{9.3}$~M$_{\odot}$
for a median SFR of $\sim$2.5~$M_{\odot}$yr$^{-1}$ which is five times higher than in the LMC.
A large fraction of GRB hosts are the equivalent of local starbursts.


\bigskip
$\bullet$ {\sl Metallicity}

\begin{figure}
\centering
\includegraphics[width=8.5cm, height=7cm, angle=0]{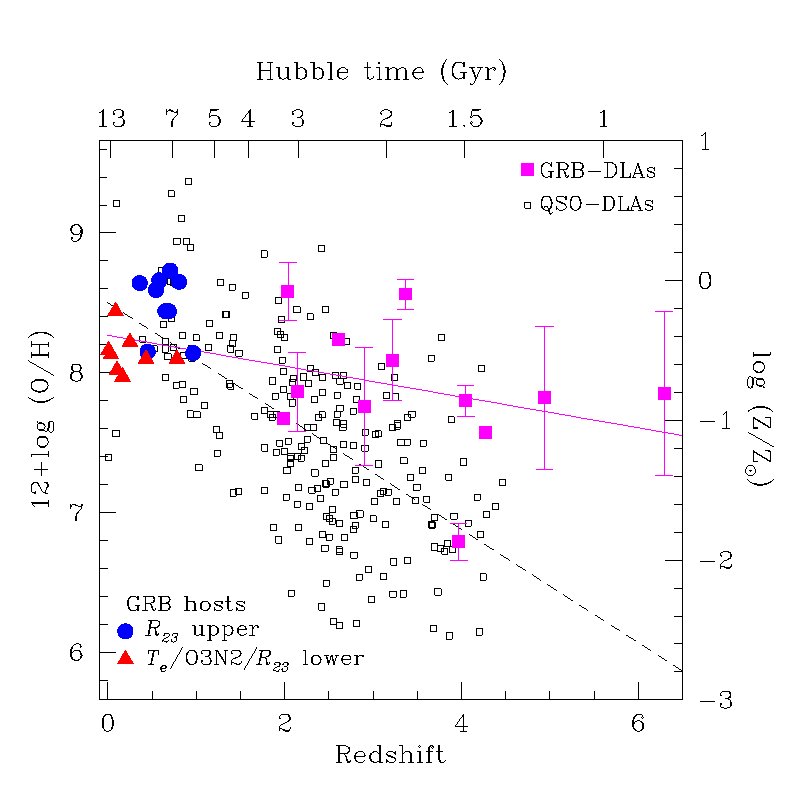}
\caption{Metallicities as a function of redshift (lower $x$-axis) or Hubble time (upper $x$-axis).
The filled blue circles are upper limits on the GRB-host metallicities (upper branch of the $R_{23}$ indicator;
the lower branch gives values about 0.5~dex less).
The filled red triangles are GRB-host metallicities from another indicator ($T_{\rm e}$ and O3N2).
Filled squares are metallicities measured in the GRB ISM from Damped Lyman-$\alpha$ systems at the redshift of the GRB
(see Section~\ref{ISM}; \cite{Savaglio2006}) and open squares are metallicities measured in QSO intervening
DLAs \cite{Prochaska2003}. The Figure is from \cite{Savaglio2009}.
}
\label{Metallicity}
\end{figure}

Metallicities in the ionized gas of the host galaxy are measured from emission lines using different
indicators. When the electronic temperature, $T_{\rm e}$, can be derived (usually from the
[OIII]$\lambda$4363/[OIII]$\lambda$5007 line ratio), then the metallicities can be calculated 
straightforwardly assuming the electronic density is smaller than $\sim$10$^{3}$~cm$^{-3}$ \cite{Izotov2006}. 
Note that measurements from the [OII] doublet show that the electronic density in GRB hosts may not be much
smaller than this limit \cite{Savaglio2009}. The difficulty with this method is that the [OIII]$\lambda$4363 
emission line is usually weak and often stays undetected. In that case, metallicities are estimated from  
the $R_{23}$~=~([OII]+[OIII])/H$\beta$ or $O3N2$~=~log([OIII]/H$\beta$)$-$log([NII]/H$\alpha$) line ratios 
(e.g. \cite{Pettini2004}). However, when the temperature is unknown, there is a degeneracy in the measurement of 
metallicity. The reason is that the forbidden lines are collisionally excited so that high metallicity and low temperature
(higher branch) is equivalent to low metallicity and high temperature (lower branch).
Results are shown in Fig.~\ref{Metallicity} where measurements for GRB hosts are plotted
versus redshift together with measurements in Damped Lyman-$\alpha$ (DLA) systems observed at the redshift  
of the GRB in the afterglow optical spectrum (see Section~\ref{ISM}; \cite{Savaglio2006}) and
measurements in intervening DLA systems observed in quasar spectra \cite{Prochaska2003}.
GRB metallicities at $z<1$, although low ($<$0.1~Solar) are consistent with what is measured
in intervening galaxies.
Recall that DLAs are very strong ($W_{\rm r}$~$>$~5~\AA) H~{\sc i} Lyman-$\alpha$ absorptions 
equivalent to what would be observed along a line-of-sight passing through the disk of our galaxy.

\bigskip
$\bullet$ {\sl More data ?}

From the study of \cite{Savaglio2009}, it seems that long GRB host-galaxies are faint, blue, low-mass and low-metallicity 
starforming galaxies detected at any redshift just because a GRB event has occurred. 
The authors conclude that host galaxies do not seem to be more metal-poor than normal star-forming galaxies with 
similar masses. 

However Levesque et al. \cite{Levesque2010} compare the long GRB host
mass-metallicity relation in a sample of (only) 8 objects to that of samples
representative of the general star-forming galaxy population, and conclude
that long GRBs occur in
host galaxies with lower metallicities than the general population, and that this trend extends to
$z\sim1$, with an average offset of $-0.42\pm0.18$ from the M-Z relation for star-forming galaxies.

In addition, \cite{Han2010} searched for the spectral signature of Wolf-Rayet stars in
the host-galaxy of eight GRBs with $z<1$ (see also \cite{Hammer2006}). They detect WR stars in 5 GRB host galaxies and show that
the GRB host galaxies have slightly lower metallicities compared to SDSS galaxies of comparable luminosity and stellar
mass. The presence of WR stars and the observed high WR/O star ratio, together with the low metallicity,  
imply that little star formation has occured yet in these galaxies.
Moreover, \cite{Stanek2006} have shown that long-duration gamma-ray bursts 
GRBs 980425, 020903, 030329, 031203 and 060218, each of which had a well-documented associated supernova, 
have all faint and metal-poor host galaxies compared to the population of local star-forming galaxies. 

Although the conclusions by \cite{Savaglio2009} seem reasonable, it is probably wise to wait for more data to be gathered 
before concluding that GRB host galaxies are drawn from the normal population of star forming galaxies
at $z<1$.
 
\subsubsection{The high-$z$ view; $z>2$}

The detection of GRB hosts at redshift $z>2$, if they are drawn from the general population
of galaxies as expected, will allow us to
examine in detail whether the mass-metallicity and luminosity-metallicity relations 
exist at those high redshifts, and, in that case, if they follow the evolutionary
trend observed at lower redshifts that galaxies of a given mass or
luminosity have lower metallicities at progressively higher redshifts. 
Moreover, with the ability to probe galactic-scale outflows in absorption from the spectra
of the afterglow, we can determine
whether the origin of these relations is rooted in higher gas fractions for lower mass galaxies
(therefore diluting the metals), or in more efficient outflows from the shallower
potential wells of low-mass galaxies (expelling the metals more easily in the
intergalactic medium; e.g. \cite{Erb2006}). Other important questions 
such as the Lyman-$\alpha$ emission from these galaxies or the production of metals
would benefit from these observations (e.g. \cite{Niino2009}\cite{Fan2010}).
The recently installed spectrograph VLT/X-shooter will boost this field in the next few years.

The to date biggest sample of host galaxies at $z>2$ has 15 objects \cite{Chen2009} observed
with the {\sl HST}. It is found that the UV luminosity distribution of GRB host galaxies is consistent 
with expectations from a UV luminosity weighted random galaxy population with a median luminosity of 
$L({\rm UV}) = 0.1 L_{\rm *}$ but the UV radiation field in the ISM must be strong.  
There is moderate evidence for the presence of outflows and tentative evidence for a trend 
of declining ISM metallicity with decreasing galaxy luminosity. GRB host galaxies at $z > 2$ (with 
known optical afterglows) are representative of unobscured star-forming galaxies.
However, very recently, GRB~080607 at $z=3.36$ has been found to be associated with a mature dusty galaxy
with SFR~=~125~$h^{-2}$M$_{\odot}$/yr and a total stellar mass of $M_{\rm *}$~$\sim$~4x10$^{11}$~$h^{-2}$M$_{\odot}$
\cite{Chen2010}. The GRB afterglow had the particularity to be highly extinguished ($A_{\rm V}\sim ~3$~mag).
Since the localization of such extinguished afterglows can be difficult,
it is possible that the current sample of host-galaxies is biased against dusty and high-metallicity galaxies.

At $z>4$ the detection of host galaxies is even more difficult and is time consuming even with the currently
largest telescopes. For example, the host galaxy of GRB~090205 at $z_{\rm GRB}$~=~4.65
was detected in the R and I-band ($m_{\rm AB}$~=~26.40, 25.22 respectively) but
not in J, H and K bands with limiting magnitudes of, respectively, 24.4, 24.2, 23.9, 
after 1~h integration time in each band with HAWK-I on VLT \cite{DAvanzo2010}.
The detection is even more difficult at higher redshift, as shown by the non-detection of the host of 
GRB~050904 at $z=6.295$ in {\sl HST} and {\sl Spitzer} data \cite{Berger2007b}.

\subsection{The host galaxies of short GRBs}

It has been found that short GRBs at low redshift originate in a variety of environments 
that differ substantially from those of long GRBs, both on individual galaxy scales and on galaxy-cluster 
scales \cite{Prochaska2006}\cite{Leibler2010}. Some have been found associated with old and massive galaxies with no current 
($<0.1$~M$_{\odot}$yr$^{-1}$) or recent star formation. Some have been found in clusters.
However, the majority of short GRBs appear to occur in star forming galaxies, raising the possibility 
that some progenitors are related to recent star formation activity. Berger \cite{Berger2009}
showed that these star-forming galaxies have
luminosities of $L_{\rm B}\sim 0.1-1.5~L^{\rm *}$, star formation rates of 
$SFR\sim 0.2-6$~M$_{\odot}$yr$^{-1}$, and metallicities of $Z \sim 0.6-1.6$~$Z_{\odot}$. 
A detailed comparison with the hosts of long GRBs reveals systematically higher luminosities, lower specific 
star formation rates (SFR/$L_{\rm B}$) by about an order of magnitude, and higher metallicities by about 
0.6~dex. The Kolmogorov-Smirnov probability that the short and long GRB hosts are drawn from the same underlying 
galaxy distribution is only $\sim$10$^{-3}$. Short GRB hosts exhibit excellent agreement with the specific star 
formation rates and the luminosity-metallicity relation of field galaxies at $z\sim 0.1-1$. 
They are not dominated by young stellar populations like long GRB hosts. Instead, short GRB hosts 
appear to be drawn uniformly from the underlying field galaxy distribution, indicating that the progenitors 
have a wide age distribution of several Gyr. 


\subsection{Dark GRBs}

For a significant proportion (25\%$-$50\%) of all well-localised GRBs, no optical/near-infrared
afterglow is detected and/or the optical afterglow emission is lower than that expected from the X-ray afterglow emission
(e.g. \cite{Fynbo2009}). These bursts are called ``dark GRB''. The nature of the dark bursts is still to be understood, although
several ideas have been proposed to explain why some bursts are dark in the optical bands: (i) shifting of the rest-frame optical
afterglow emission and the Lyman-limit towards longer wavelengths for high redshift bursts, (ii) intrinsic dimness of the afterglow,
or (iii) high extinction in either the host galaxy or the circumburst environment along the line of sight \cite{Greiner2011}.

Host galaxy studies of dark GRBs indicate that the majority of dark GRB hosts are similar
to normal long GRB hosts, which do not have very high extinction or very high redshifts \cite{Perley2009}. 
However, a few dark GRBs have been observed to have both high extinction and near-solar metallicities derived from 
afterglow observations (e.g. \cite{Kruhler2008}\cite{Prochaska2009}\cite{Kupcu2010}\cite{Holland2010}), a few of them 
also exhibiting high
SFR. It is possible that such hosts are more common than previously anticipated and 
that a fraction of dark GRBs explode in regions dusty enough so that their afterglow is very hard to detect in the optical.

\section{The Interstellar Medium of GRB host galaxies}
\label{ISM}

The spectra of high redshift GRB afterglows show a welth of absorption lines at the redshift of the GRB, the
most striking feature being a very strong, most of the time damped, Lyman-$\alpha$ H~{\sc i} absorption 
produced by a large column density (log~$N$~$>$~20.3) of neutral gas located in the ISM of the host-galaxy. 

The properties of the interstellar medium (ISM) of high redshift galaxies, its metallicities 
and physical state, are usually derived from the observations of intervening Damped Lyman-$\alpha$ systems
along the line of sight to bright quasars (e.g. \cite{Wolfe2003}\cite{Noterdaeme2009}).
However, such measurements are likely biased toward regions of low density (\cite{Rauch2008};
see however the efforts to detect molecules in DLAs \cite{Noterdaeme2008}\cite{Srianand2008}).
On the contrary, long GRBs are probably associated with the explosion of massive stars and are therefore
likely to be associated with the very regions where stars form. They are therefore unique
probes of the environment of high density gas and molecular clouds (e.g. \cite{Campana2007}\cite{Prochaska2009}). 
Note that, since intervening and GRB DLAs do not trace the same regions, 
we should expect differences between the properties of these two populations.
 
What makes GRBs special is that it is in principle possible (if good and suitable data are taken) to 
observe both the absorptions due to the ISM of
the GRB host-galaxy and the emission (continuum from stars and emission lines from the 
ionized gas) from the host itself. Unfortunately, as seen previously, galaxies are difficult to
detect and study at $z>2$ whereas absorption lines are more easily observed at these redshifts.
The latter is due to the fact that most of the interesting and useful absorption lines have rest-wavelengths in
the UV and are redshifted in the optical for $z\geq2$. 

Since log~$N$(H~{\sc i})(cm$^{-2}$) is so high ($>$~20) in DLA systems, 
(i) it is possible to detect species of low abundances and, of particular interest in the 
case of GRBs, absorption lines from excited levels of atomic ground states; (ii) metallicities are easily derived
because, at such high $N$(H~{\sc i}), most of the species are in their neutral or singly ionized stages
and metallicities can be derived with high accuracy from log~$N$(X)~$-$~log~$N$(H~{\sc i}) 
(where X~=~O~{\sc i}, S~{\sc ii}, Fe~{\sc ii}, Si~{\sc ii}, etc...) without any ionization correction. 
One should be aware however that if little error is to be made on $N$(H~{\sc i}) from the fit
of the damped wings of the Lyman-$\alpha$ absorption line, errors on $N$(X) can be quite large
if (i) absorption lines are saturated and (ii) the spectrum is not of high enough spectral resolution 
This should be remembered before embarking in discussions based on inappropriate results 
(see e.g. \cite{Jenkins1986}\cite{Prochaska2006b}).

Spectra should be of high quality for this kind of studies, therefore they should be taken very shortly
after the gamma burst, when the afterglow is still bright. At the European Southern Observatory's (ESO) Very Large
Telescope (VLT) and at other telescopes in the world, a Rapid Response Mode (RRM) has 
been commissioned to provide prompt follow-up of transient phenomena. The design of this system 
allows for completely automatic observations without any human intervention except for 
the placement of the spectrograph entrance slit on the GRB afterglow. The typical time 
delay on VLT, which is mainly caused by the telescope preset and object acquisition, is 5$-$10~min
\cite{Paul2007}. 
 
\subsection{Variable absorption lines}

The GRB afterglow radiation is  intense enough to have an important impact on its 
environment at the time of the explosion. In particular, the UV radiation
will ionize the neutral gas (e.g. \cite{Perna1998}) and destroy molecules and dust grains
up to tens of parsecs away \cite{Waxman2000}\cite{Draine2002}.
Interestingly, rotational levels of molecules and metastable states of
existing species (O~{\sc i}, Si~{\sc ii}, Fe~{\sc ii}) are populated by UV pumping followed
by radiative cascade. After these levels are populated, they can induce
corresponding absorption lines in the afterglow spectrum (e.g. \cite{Paul2004}). 
Note that excitation of metastable states 
by fluorescence of UV light or direct excitation by IR radiation has been
observed in a few quasars (see e.g. \cite{Wampler1995}\cite{Srianand2000}).

In addition, since the GRB afterglow fades rapidly, recombination prevails and the populations of these levels
change (most of the time decrease but sometimes can increase for a while). This induces 
these absorptions in the afterglow spectrum to vary till to disappear.
Detection of these time-dependent processes, with timescales ranging from seconds 
to days in the observer frame can lead very interesting information on the 
burst itself and the ISM of the host.

These phenomena can be observed if several spectra are taken within a few minutes from the burst
\cite{Dessauges2006}\cite{Paul2007}. Five epochs from 10 to 80 minutes 
after trigger have been used by \cite{Paul2007} to model the time variation of Fe~{\sc ii} and
Ni~{\sc ii} absorption lines in UVES spectra of the GRB~060418 afterglow. They find strong evidence 
for the UV excitation mechanism. The striking result of their study however is that
most of the gas is located at a distance  of about 1.7$\pm$0.2~kpc from the GRB. This distance
is found to be 440$\pm$30~pc in the case of GRB~050730 \cite{Ledoux2009}. These distances
are surprisingly large since most GRBs are expected to be associated
with the very regions where stars form, therefore gas is expected to be present and detectable
much closer to the GRB.
On the other hand, the detection of Mg~{\sc i} absorptions from the ISM gas implies  
that the neutral gas should be located at distances larger than $\sim$50~pc from the GRB afterglow
\cite{Prochaska2006c}. 
  
Note that it has been argued  from X-ray data
for a time-variable absorbing column density \cite{Starling2005}\cite{Campana2007}, presumably due to 
the ionization of the gas. Much of the X-ray absorbing gas is situated very close to the GRB, whilst the 
H~{\sc i} absorption causing the DLA is most likely located further out (e.g. \cite{Schady2011}).

\subsection{The GRB DLAs}

It is not surprising to observe a strong H~{\sc i} Lyman-$\alpha$ absorption (DLA) at the redshift
of the GRB since, the GRB is expected to be associated with
a star forming region where the gas is likely to be found in large quantities. 

DLAs are well studied along the line of sight to quasars (QSOs). Although it is clear that DLAs are located close 
to regions where stars form (presence of metals, detection of C~{\sc ii}$^*$ absorption, galaxy-like kinematics;
\cite{Wolfe2003}), their exact nature is not completely elucidated. Some should be associated with disks of 
galaxies especially when molecules are detected (in $\sim$10~\% of the DLAs; \cite{Noterdaeme2008}),
others are probably located in the outskirts of galactic haloes \cite{Fox2007}\cite{Rauch2008}.
 
The difference between the two populations of GRB and QSO DLAs is apparent because 
(i) a significant number of GRB DLAs have H~{\sc i} column densities
well beyond 10$^{22}$~cm$^{-2}$ \cite{Paul2004}\cite{Savaglio2006}\cite{Fynbo2009}\cite{Savaglio2010} 
when there is only one QSO-DLA with
log~$N$(H~{\sc i})~$>$~22 in the whole SDSS \cite{Noterdaeme2009} and (ii) the mean metallicity is
larger for GRB-DLA (see Fig.~\ref{Metallicity} and \cite{Savaglio2004}) with a mean metallicity of about 0.1~solar 
at $z>2$ to be compared to $\sim$0.03~solar for intervening DLAs. 
However, it should be noted that (i) low metallicity GRB-DLA do exist ($Z$~$\sim$~10$^{-2}$$Z_{\odot}$; \cite{Rau2010})
although systematics may underestimate most of the measurements \cite{Prochaska2006b}; and
(ii) the vast majority of QSO-DLAs have metallicities well above 10$^{-3}$$Z_{\odot}$ (see Fig.~\ref{Metallicity}
and \cite{Penprase2010}). 

Molecules (H$_2$, CO) have been detected in different QSO-DLAs \cite{Noterdaeme2008}, whereas they are rarely
seen in GRB-DLAs \cite{Ledoux2009}. This may be a consequence of small statistics (see the only detection in 
\cite{Prochaska2009}). In addition, the sample of GRBs with optical afterglow spectroscopy is not representative of all Swift bursts, 
most likely due to a bias against the most dusty sight lines \cite{Fynbo2009}. Here again, the recently 
installed VLT spectrograph X-shooter should help. Note however that reliable detailed studies 
of the ISM physical state can only be performed with high spectral resolution spectrographs like UVES 
(see \cite{Noterdaeme2010}).

Some cosmological simulations show that the host of GRB-DLAs are predominantly associated with dark matter haloes of mass 
10$^{10}$~$<$~$M_{\rm vir}$/$M_{\odot}$~$<$~10$^{12}$, an order of magnitude larger than those of
QSO-DLA hosts (\cite{Pontzen2010}; see however \cite{Barnes2010}) and that
the median impact parameters between the line of sight
toward a GRB or a QSO and the center of the closest galaxy may be approximately 1~kpc for 
GRB-DLAs (at the redshift of the GRB) compared with 4~kpc for intervening DLAs toward quasars.
We wonder however if this difference is meaningful
and suspect the resolution of
the current simulations (both spatial and in the description of the adequate
physical processes) may not be high enough for such an analysis. 


\section{Intervening absorbers along GRB lines of sight}

\begin{figure}[htp]
\begin{center}
\includegraphics[width=8cm]{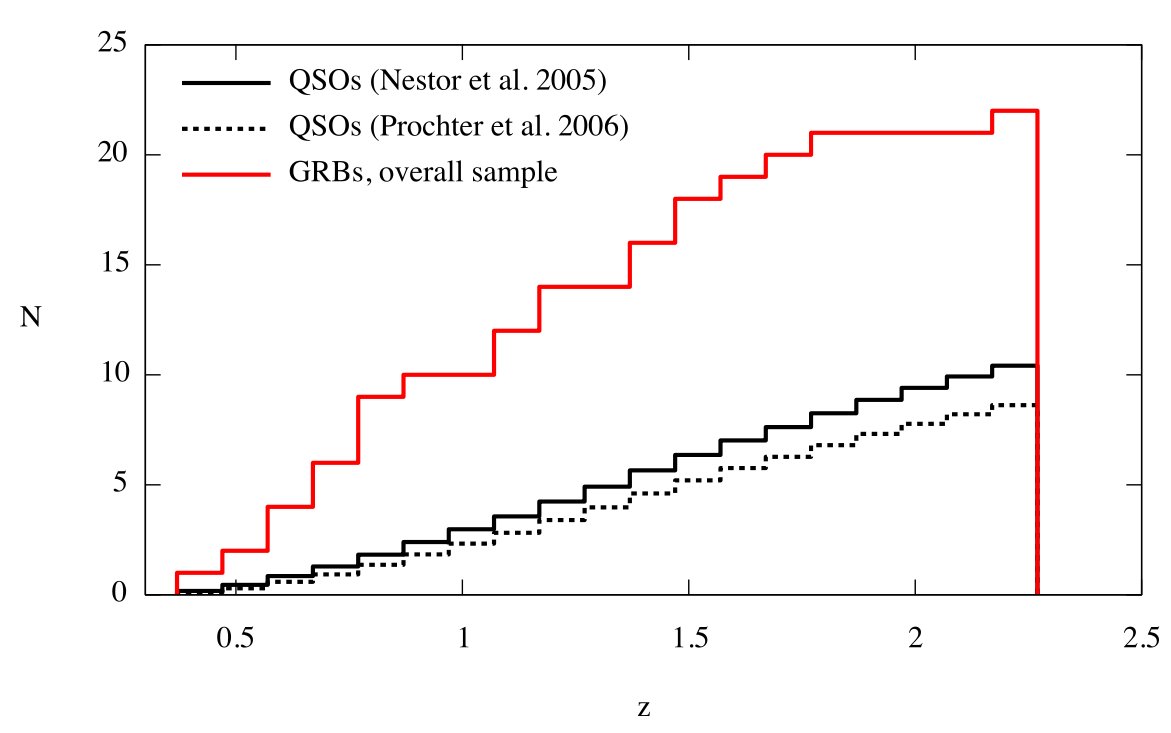}
\caption{Comparison between the cumulative distribution of strong Mg~{\sc ii} systems ($W_{\rm r}$~$>$~1~\AA) 
in the  sample of GRB lines of sight (red upper histogram) and the one expected along QSO line of sight  
(black solid and black dashed histograms). The number of strong Mg~{\sc ii} systems is significantly
larger in front of GRBs. The Figure is from \cite{Vergani2009}.}
\label{MgII}
\end{center}
\end{figure}
It was claimed by \cite{Prochter2006} that the number of strong (rest equivalent width $W_{\rm r}>1$\,\AA) 
intervening Mg~{\sc ii} absorbers is more than 4 times larger along GRB lines of sight than what is measured for 
QSOs over the same path length. 
Dust extinction bias along QSO lines of sight, gravitational lensing, contamination 
from high-velocity systems local to the GRB (see also \cite{Bergeron2011}) and differences in beam sizes are among the possible 
causes of this discrepancy, but no convincing explanation has been found yet \cite{Porciani2007}. 
Although their number density has been confirmed to be about
twice larger in front of GRBs than in front of QSOs (\cite{Vergani2009}; see Fig.~\ref{MgII}), both the estimated dust extinction and the 
equivalent width distributions of strong Mg~{\sc ii} systems along GRB lines of sight are 
consistent with what is observed in standard DLAs.
This, together with the fact that the number of weak Mg~{\sc ii} systems are not different,
supports the idea that current sample of GRB lines of sight
could be biased by a subtle gravitational lensing effect \cite{Vergani2009}.
It is therefore of first importance to search the GRB fields for galaxies associated with these 
intervening Mg~{\sc ii} systems to understand this excess. The study of these Mg~{\sc ii}-galaxy pairs 
is in addition interesting by itself as it is a unique opportunity to 
reveal directly the exchanges of gas between the IGM and galaxies (see \cite{Pollack2009}
and also \cite{Churchill2007} and references therein for QSO studies).

\section{GRBs and Cosmology}

\subsection{The highest redshift objects} 
GRB090423, at a redshift of $z=8.26^{+0.07}_{-0.08}$ \cite{Tanvir2009}\cite{Salvaterra2009}, 
is the object with the second highest redshift known to date after
the recently discovered galaxy at $z=8.5$ \cite{Lehnert2010}, to be compared with the highest
redshift quasar at $z=6.43$ \cite{Willott2003}. 
It establishes that massive stars were being produced, and dying as GRBs, $\sim$625~million 
years after the Big Bang. GRBs will be therefore 
unique probes of the period over which the intergalactic medium has been reionized by the ultra-violet 
radiation emitted by the first generations of massive stars \cite{Faucher2008}. This period is believed to 
extend from $z=17$ down to approximately $z=6$ for hydrogen and $z=3.5$ for helium, depending 
on how it proceeds.
GRBs will help understand how the universe has been reionized but also
how the intergalactic medium has been enriched with metals, what is the nature of the first collapsed objects in the Universe and related questions \cite{Quinn2010}\cite{Gallerani2008}.

It is very important to search in the spectrum of high redshift ($z>6$) GRB afterglows for the red damping wing of the absorption trough produced by neutral hydrogen
in the IGM. As it is known that reionization ends up around $z\sim6$ \cite{Becker2007},
this signature would be definite evidence for this crucial phase of the universe history. 
GRBs have the advantage that no large scale proximity effect is expected contrary to QSOs that ionize
the IGM to a distance of several Mpc \cite{Guimaraes2007}. Their disadvantage is that
usually neutral gas from the host galaxy located 
in front of the GRB already produces a damping wing which is difficult to 
disentangle from the IGM effect \cite{Totani2006}\cite{Patel2010}. 

\subsection{GRBs as standard candles ?} 
Among GRBs with known redshifts, 45\% are at $z > 2$
and 8\% at $z > 4$. Their high luminosity (10$^{51}$~erg) and their detection in the $\gamma$-ray
bands make them attractive as a potential and complementary cosmological tool to
constrain the cosmological models at $z > 2$. However, GRBs are
all but standard candles \cite{Bloom2003}: their isotropic equivalent energetics and luminosities span
three to four orders of magnitudes. Similarly to SNIa, it has been proposed to use correlations
between various properties of the prompt emission \cite{Firmani2006} and of the afterglow
emission \cite{Ghirlanda2004}\cite{Schaefer2007} to standardize GRB energetics.
The most promizing correlations are those using only the temporal and spectral properties of the prompt emission
\cite{Firmani2009}. One of them is the correlation found between the
rest-frame peak energy and the isotropic equivalent energy ($E_{\rm peak}-E_{\gamma}$ correlation;
\cite{Ghirlanda2009}\cite{Amati2010}). However even this correlation is debated due to possible selection effects and 
to controversial estimates of $E_{\gamma}$.
Thus, for the moment, it may be hazardous to derive any constraint on cosmological parameters
using GRB properties. It is worth however pursuing in a direction that may well reveal itself
of invaluable reward in the future.

\section{Conclusion}

GRBs are unique probes of the distant universe. Long GRBs are probably associated with the 
explosion of massive stars and therefore reveal at very high redshift the death of the first stars. 
They will be in the future unique probes of the period over which the intergalactic medium 
has been reionized by the ultra-violet radiation emitted by the first generations of massive 
stars and the first collapsed structures. GRBs are somehow correlated with
the activity of star-formation and could be used to follow the SFR to the highest redshift.
They are likely to be associated with the very regions where stars form and are unique
probes of the environment of high density gas and molecular clouds 

GRBs flag remote galaxies that would probably stay unnoticed otherwise.
GRB host galaxy studies carried out up to now show that
long GRB host-galaxies are faint, blue, low-mass and low-metallicity 
starforming galaxies detected at any redshift just because a GRB event has occurred. 
The short and long GRB hosts are drawn from galaxy populations that are statistically 
different, with the short GRB hosts having higher metallicity and higher luminosity
(for the same SFR).
It is however possible that the current sample of host-galaxies is biased against
dusty and high-metallicity galaxies.

What makes GRBs special is that it is in principle possible (if good and suitable data are taken) to 
observe both (i) the absorptions due to the gas associated with the galaxy, either the disk, or  
outflows to or inflows from the IGM, or the outskirts of the galactic halo, that are detected
in the spectrum of the afterglow and (ii) the emission (continuum from stars and emission lines from the 
ionized gas) from the host itself. This can yield unique information on the
exchanges between the galaxy and the intergalactic medium that are crucial to explain how
galaxies form and evolve.

In the near future, new capabilities such as X-shooter will make it possible
to increase the sample of well observed afterglows and host-galaxies.
The french-chinese gamma-ray satellite {\it SVOM}, which launch is foreseen for 2015,
has been optimized to increase the number of high redshift GRB detections and the synergy with the ground-based 
facilities in order to favour the rapid follow-up of the afterglow.
This will boost the amount of information available to tackle the important
issues revealed by this exciting field of research.

 



\end{document}